\documentclass[a4paper]{article}
\usepackage{verbatimbox}
\usepackage{INTERSPEECH2021}
\usepackage{color}
\usepackage{ulem}
\usepackage{subfigure}

\title{A Study into Pre-training Strategies for Spoken Language Understanding on Dysarthric Speech}
\name{Pu Wang$^1$, Bagher BabaAli$^2$, Hugo Van hamme$^1$}
\address{
  $^1$Department of Electrical Engineering-ESAT, KU Leuven, Belgium\\
  $^2$School of Mathematics, Statistics and Computer Science, College of Science, University of Tehran, Iran}
\email{pu.wang@esat.kuleuven.be, babaali@ut.ac.ir, hugo.vanhamme@esat.kuleuven.be}

\begin{document}

\maketitle
\begin{abstract}
  End-to-end (E2E) spoken language understanding (SLU) systems avoid an intermediate textual representation by mapping speech directly into intents with slot values. This approach requires considerable domain-specific training data. In low-resource scenarios this is a major concern, e.g., in the present study dealing with SLU for dysarthric speech. Pretraining part of the SLU model for automatic speech recognition targets helps but no research has shown to which extent SLU on dysarthric speech benefits from knowledge transferred from other dysarthric speech tasks. This paper investigates the efficiency of pre-training strategies for SLU tasks on dysarthric speech. The designed SLU system consists of a TDNN acoustic model for feature encoding and a capsule network for intent and slot decoding. The acoustic model is pre-trained in two stages: initialization with a corpus of normal speech and finetuning on a mixture of dysarthric and normal speech. By introducing the intelligibility score as a metric of the impairment severity, this paper quantitatively analyzes the relation between generalization and pathology severity for dysarthric speech.
\end{abstract}

\section{Introduction}
A spoken language understanding (SLU) system that converts speech into the desired intent and/or a set of actions is a crucial part for spoken user interfaces like personal assistants and home automation agents \cite{1}. A traditional SLU system is implemented as a pipeline which consists of an automatic speech recognition (ASR) module followed by a natural language understanding (NLU) module and suffers from error propagation generated by the separate training of these two modules \cite{2}. End-to-end (E2E) SLU addresses this \cite{3}, by directly mapping speech to the pre-defined semantic slots, e.g., a command “turn on the light in the bedroom” would be directly mapped to an intent and associated slot values: “\textit{\{action: switch on, location: bedroom, object: light\}}” without an intermediate text representation.
  
However, it is believed that E2E SLU is data-hungry since it is typically constructed as a deep learning approach requiring in-domain training data to achieve good generalization across the differing linguistic habits of its users. Therefore, SLU tasks in low resource languages or domains usually fail to achieve a performance comparable to what we are acquainted with today in e.g., state-of-the-art personal assistants in English. Even stronger problems are experienced by speakers with a voice pathology because this type of speech is usually not included in the training and the broad umbrella encompassed by “speech disorders”. Training data is very scarce for several reasons including the increased effort required to collect data from this population \cite{4,5}.

\cite{6,7,8} present a remedy for low resource SLU by designing a user-taught SLU system. “User-taught” refers to the strategy in which the SLU agent learns from scratch only based on spoken commands and corresponding task demonstrations from its users. This implies the system becomes speaker-dependent and does not need to make any assumptions on the diverse grammatical and lexical preference of its users. Therefore, its requirement on training data size is naturally lower. However, users will have to provide the training samples themselves. To ensure the involved user effort is moderate, the designed SLU system typically has a compact and highly efficient structure that can quickly converge after only a few training samples.
  
Another solution to deal with the data scarcity issue could be pre-training \cite{3,9,10,11,12}. In our past work, we have shown the combination of pre-training strategies and the user-taught SLU helps with normal speech \cite{12}. We know of no previous work to investigate whether pre-training on dysarthric speech would benefit the dysarthric SLU task, while such studies are available for the ASR task \cite{13, 14}. An essential difference is that in user-taught SLU, the E2E training might learn to correct ASR errors, reducing benefits of pretraining.
\begin{figure*}[t]
  \centering
  \includegraphics[width=0.95\linewidth]{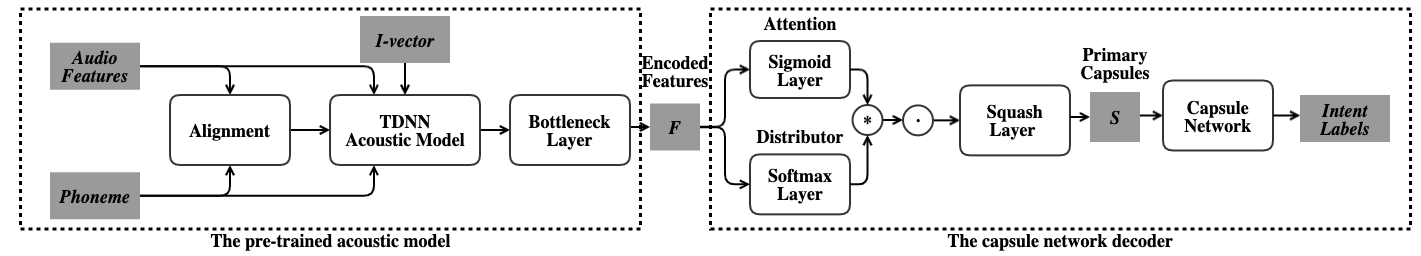}
  \caption{Structure of the dysarthric SLU system.}
  \label{fig1}
\end{figure*}  

In this paper, the implementation of the aforementioned user-taught SLU system is adapted to focus on dysarthric speech. Previous studies have demonstrated bottleneck features (BNF) yield stable representations of dysarthric speech \cite{13,14}, while \cite{15} shows improvements in dysarthric ASR by exploring a time-delay neural network (TDNN). Inspired by these, we explore to pre-train a TDNN based acoustic model on a mix of normal and dysarthric speech with ASR targets and extract layer activations of the TDNN model as BNFs. The extracted speech encodings are fed to a 2-layer capsule network decoder to achieve the E2E SLU task. Since the acoustic model is not likely to learn well from dysarthric speech directly, we utilize a two-stage training. The acoustic model is initially pre-trained on a corpus of normal speech and then fine-tuned on a mixture of dysarthric and normal speech. As speech impairments also differ with the severity of the disorder, \cite{16} attempts to utilize the percentage of consonant correct (PCC) index to diagnose the disordered severity and reveals a potential relation between severity-level and speaker adaptation in the ASR task. We furthermore introduce the speaker’s intelligibility score (IS) as a metric of the severity of impairment severity for both the pre-training corpus and the target utterances. By comparing ASR results and SLU accuracies of the multiple acoustic models pre-trained on impaired utterances collected from different IS ranges, we quantitatively analyze opportunities for adaptation towards pathological speech. The contribution of this paper is hence:

1) presenting a usable implementation of knowledge transfer for the acoustic model pre-trained on the dysarthric speech to the dysarthric SLU task.

2) quantitatively analyze the relation between pathological speaker adaption and impairment severity.
%\begin{enumerate}
%\item presenting a useable implementation of knowledge transfer for the %acoustic model pre-trained on the dysarthric speech to the dysarthric SLU %task.
%\item quantitatively analyze the relation between pathological speaker %adaption and impairment severity.
%\end{enumerate}

In section 2 we detail the pre-trained acoustic model, pre-training material as well as the two-stage training strategy. The overall structure of the designed dysarthric SLU system and capsule network decoder are introduced in this section as well. Section 3 discusses the specific experimental setting for evaluation, and the corresponding results will be presented in section 4. In section 5 we will conclude our work.

\section{Model}
The overall structure of the dysarthric SLU system consists of the pre-trained acoustic model and a capsule network decoder, as shown in the Figure 1. The acoustic model is pre-trained on a dysarthric speech corpus as well as on a larger-scale corpus of normal speech. After training, the acoustic model will be frozen. In the SLU task, the pre-trained acoustic model is utilized to extract the high-level BNFs of the input speech from its final network layer. The BNFs will be decoded by the 2-layer capsule network to yield the intent labels and slot values.

\subsection{Pre-trained ASR acoustic model}
\subsubsection{ASR acoustic model}
The ASR model is built with the Kaldi software \cite{17}. We use the HMM-GMM model to first align audio features with a sequence of context-dependent phonemes. The obtained phoneme-audio alignments are then used to train the 16-layer TDNN acoustic model with 1536 dimensions for each layer. We use 40-dimensional MFCC features for acoustic model training. To learn deviations of intra-speakers \cite{10}, 100-dimensional I-vectors are appended to the MFCC features before feeding them into the TDNN model.

After pre-training the acoustic model, the model’s parameters are frozen during the SLU task. For each utterance in the SLU task, we extract its 160-dimensional BNFs from the 16th TDNN layer as the encoded features and feed them to the intent decoder for the E2E SLU task training.
\begin{table}[ht]
  \setlength{\abovecaptionskip}{12pt}
  \caption{Statistics of the Copas corpus}
  \label{tab:table1}
  \centering
  \addvbuffer[-0pt -24pt]{
  \begin{tabular}{ccc}
    \toprule
    \textbf{Severity(IS)}& \textbf{\# of speakers}& \textbf{\# of hours}\\
    \midrule
    Mild ($>85$) & $99$ & $1.95$\\
    Moderate ($70$--$85$) & $63$ & $1.41$\\
    High ($60$--$70$) & $8$ & $0.2$\\
    Severe ($<60$) & $12$ & $0.4$\\
    Total & $182$ & $3.96$\\
    \bottomrule
  \end{tabular}}
\end{table}
%\vspace{-1cm}
\subsubsection{Pre-training data}
The pre-training data originate from two corpora.

Corpus Gesproken Nederlands (CGN) \cite{18} is a corpus of normal Dutch speech as spoken in Flanders and The Netherlands. We include all data from the Flemish part except the narrow-band recordings (corpus components c and d) and the spontaneous conversations (component a). The training data is composed of 138297 utterances containing 76115 word forms, which is about 133 hours in total.

Copas is a Dutch corpus of pathological speech recorded in Flanders \cite{19}, including data for Dutch intelligibility assessment (DIA) for each speaker. It contains 10792 utterances with 1160 word forms, of which 575 words occur in CGN. We summarize the speaker information and IS in Table 1. The IS is the percentage of correctly perceived phonemes by experienced speech-language-pathologists. The speech recordings are divided into 4 severity levels based on these scores. The highest IS is 100, i.e., normal speech, and the lowest score is 28 which is considered as severely impaired.

\subsubsection{Pre-training strategy}
Pre-training is organized in two stages. In the first stage, we build an initial acoustic model on the normal CGN corpus to map to 218 context-dependent phone states. To improve the robustness, the CGN corpus is augmented with speed perturbation with ratio 0.9, 1.0 and 1.1. The initial training takes 2 days on a single GPU for 182 iterations with learning rate varying from 0.00025 to 0.000025. The initial acoustic model will serve as the baseline to analyze the efficiency of pre-training with the dysarthric speech corpus.

In the second stage, we fine-tune the initial acoustic model with the dysarthric Copas data. To prevent the finetuned model from forgetting knowledge learned from the normal speech, we combine 4.86 hours speech from CGN with Copas to conduct the joint training during finetuning. The combined fine-tune data is augmented with the same speed perturbation as the former stage to triple the training samples. The finetune training takes around 2 hours for 14 iterations with learning rate varying from 0.00025 to 0.000025 as well. In both pre-training stages, we utilize the HMM-GMM model to generate phoneme-audio alignments before updating the TDNN acoustic model.

As argued in the introduction, training on low IS data may degrade the acoustic model due to strong deviations in pronunciation and timing. To further investigate the influence of pre-training with data from different dysarthria severity levels, in the finetuning stage, we combine the full Copas data, Copas data with IS $>$ 60, and Copas data with IS $>$ 70 with the same part of the CGN data respectively to conduct the joint training and build three finetuned models. We will compare the results of the SLU task with BNFs extracted from these three models in the experiment section.

\subsection{Capsule network decoder}
The intent decoder is a 2-layer capsule network \cite{20}. There are 32 hidden capsules with 64 dimensions in the primary capsule layer and one output capsule for each output label with 8 dimensions in the output capsule layer. The detailed structure of the capsule network can be found in \cite{6, 20}. In general, a capsule activation is characterized by a vector with length between 0 and 1 which represents the probability of the capsule’s (presented label) occurrence, and its orientation containing latent information of the capsule.

Referring to Fig 1, the encoded features $F$ extracted from the pre-trained acoustic module are converted to primary capsule vectors $S_i$ by an attention and distributor mechanism:
\begin{equation}
  \setlength\abovedisplayskip{6pt}
  \setlength\belowdisplayskip{6pt}
  S_i = Squash(w_s\cdot\sum_t \alpha_t \delta_t{}_i F_t)
  \label{eq1}
\end{equation}

Here, $S_i$ is the vector activation for capsule $i$. $Squash()$ is soft normalization function in capsule network to ensure the length of $S_i$ lies between 0 and 1. $w_s$ are trainable weights of the squash layer. $\alpha_t$ is the attention weight for each time step, which is used to filter out the unimportant time frames in the sequence (e.g., silence). $\delta_t{}_i$ are the distribution weights of distributor to assign each time step $t$ to the hidden capsule $i$.

$\alpha_t$ and $\delta_t$ are calculated from:
\begin{equation}
  \setlength\abovedisplayskip{6pt}
  \setlength\belowdisplayskip{6pt}
  \alpha_t = sigmoid(w_a\cdot F_t+b_a)
  \label{eq2}
\end{equation}
\begin{equation}
  \setlength\abovedisplayskip{0pt}
  \setlength\belowdisplayskip{6pt}
  \delta_t = softmax(w_d\cdot F_t+b_d)
  \label{eq3}
\end{equation}

Here, $w_a$ and $b_a$, $w_d$ and $b_d$ are weights and biases of the sigmoid and softmax layers respectively. A second capsule layer maps $S_i$ to the intent and slot value labels via dynamic routing \cite{20}. Essentially, the second layer learns which acoustic evidence that triggered the first layer can be pieced together as evidence for an intent or slot value.

\section{Experiments}
\subsection{Task-specific dataset}
The speech used for SLU task is from the Domotica database \cite{21}. It contains Dutch dysarthric speech commands related to home automation. Commands can be encoded in 27 slot values. A typical utterance is “turn on the kitchen light”. The corpus is recorded by 17 speakers at different times. It composed of 4174 utterances of 38 words, in which 36 words are covered by CGN and 15 words are covered by Copas. We list the severity levels based on automatically derived IS for each speaker in Table 2 (except for two children, speaker 31 and 37, for whom the automatic model does not work).
\subsection{Baseline}
The dysarthric SLU system is verified in two aspects: 1) whether the dysarthric SLU task can benefit from pre-training with ASR targets on a corpus of dysarthric speech; 2) the influence of pre-training with data from different dysarthric severity levels. Therefore, comparison experiments will be conducted on five models:
\begin{table}[hb]
  \setlength{\abovecaptionskip}{12pt}
  \setlength{\belowcaptionskip}{6pt}
  \caption{Task-specific corpus (Domotica) statistics}
  \label{tab:table2}
  \centering
  \addvbuffer[-0pt -24pt]{
  \begin{tabular}{cc}
    \toprule
    \textbf{Severity(IS)}& \textbf{Speaker IDs}\\
    \midrule
    Mild ($>85$) & $17$, $40$, $43$, $44$, $48$\\
    Moderate ($70$--$85$) & $28$, $29$, $34$, $35$, $46$, $47$\\
    High ($60$--$70$) & $30$, $32$, $33$, $41$\\
    \bottomrule
  \end{tabular}}
\end{table}

\textbf{Without pre-training:} The baseline without pre-training model is from \cite{6}. It is a state-of-art user-taught SLU system constructed as a 2-layer GRU encoder and a 2-layer capsule network decoder. This model also runs under the speaker-dependent setting and gets convincing results dealing with dysarthric speech. However, as explained in \cite{6}, the model fails to obtain the desired performance levels when it comes to limited training samples, such as less than 70 samples.

\textbf{Pre-train on CGN:} The initial acoustic model only trained on CGN corpus of normal speech is denoted as “Pre-train on CGN” model.

\textbf{Finetune on full Copas, Finetune on IS \textgreater 60,  Finetune on IS \textgreater 70}: The three acoustic models finetuned from the initial “Pre-train on CGN” model with the full Copas data, Copas data with IS $> 60$, and Copas data with IS $> 70$ respectively.

\subsection{Experimental setup}
We first compare the learning curves of the accuracy for intent label classification under the speaker-dependent setting. The learning curve records the model’s performance on an increasing amount of training data and tests on all remaining data using 5-fold cross-validation for each speaker. In each fold, the utterances from each speaker are randomly shuffled and are divided into 15 blocks \cite{6}. We increase the amount of training data from one to 14 blocks and test the model on all remaining blocks. The final learning curve is the average accuracy across all speakers.

We secondly simulate the insufficient training data situation and compare the accuracy for each speaker with features extracted from the four pre-trained acoustic models. For each speaker, we randomly select around 15\% of data as the train set, in which each command type recorded by the speaker occurs twice, the remaining samples serve as the test set. Since not all speakers record the full 27 commands, the size of the train set for each speaker varies from 18 to 54 utterances. We conduct 5-fold validation, and the final result is the average of the five results.

\section{Results}
In Figure 2, the smoothed average accuracy (the micro-averaged F1-score of detected slot values) of different models are plotted as a function of the number of training samples available for all intents. Comparing the learning curves “without pre-training” and “pre-train on CGN”, it is clear, after involving pre-training, the performances improve by up to 10\% points, which shows that the SLU task on dysarthric speech can benefit from the pre-training, even using only normal speech. Moreover, it is remarkable that with finetuning on only 3.96 hours of dysarthric speech, the accuracy increases further by up to about 5\% points with extremely small task-specific training data (at the very left of the learning curve). The results illustrate that knowledge transfer from a normal and dysarthric speech ASR task to the SLU task is possible. On the other hand, a user-taught SLU system requires user effort, therefore, an important evaluation criterion is the amount of training samples required for a given accuracy. %In Figure 2, if we consider 0.8 as the desired accuracy, without pre-training strategy, a user-taught system would require an extra 50 utterances to achieve the goal. % HVH: 0.8 is a bit of a special operating point, as you reach it with 10 utterances or so with Copas
For instance, a system without pre-training would require about 40 additional demonstrations to reach 95\% accuracy compared to pre-training on Copas. 

As aforementioned, it is hard to get accurate alignments for speech recorded from severely impaired speakers, which may cause adverse effects on the SLU task. We therefore conduct the ASR task on the Domotica dataset with acoustic model “pre-train on CGN”, and “finetune on full Copas” to figure out whether the knowledge learned from different dysarthric speakers could transfer to other speakers within different IS ranges. The word error rate (WER) is used as the evaluation criterion and results are shown in Figure 3 (a). The numbers listed near the symbols are the corresponding speaker IDs.

From Figure 3 (a), speakers with IS in the 60-75 range tend to get higher gains when the model is trained with the full Copas data. The IS of the Copas data varies from 28 to 100, and 10\% of the speech is recorded as severely impaired speech (IS below 60). With increasing IS, utterances are closer to normal speech. For instance, speakers with an IS above or close to 90 (speaker 17 and speaker 43) can be regarded to produce almost normal speech. Therefore, the improvements for speakers with IS from 75 to 85 are limited. Even worse, an adverse effect occurs with IS above 88. Inspired by the results shown in the ASR task, we further compare SLU accuracy results of four pre-train models. As we explained in the section 2.1.3, the training data of each model are collected from different IS ranges.
%, in which, “pre-train on CGN” can be seen as pre-train on utterances with IS = 100, and “finetune on full Copas” can be seen as pre-train with IS $>28$. 
The accuracy results for each speaker with “pre-train on CGN” are shown in Figure 3 (b). This result serves as the baseline. We consequently show the relative improvements against this baseline with the three finetuned models in Figure 3 (c). As demonstrated by Figure 3 (b), the pre-training strategy performs well with very limited training samples, especially for speakers with high IS, e.g., over 85. Pre-training on normal speech provides enough knowledge for utterances with high IS and therefore we should not expect a significant performance gain after involving the knowledge from dysarthric speech since their accuracies are relatively high (e.g., speaker 17 and 43). In general, the SLU accuracies show similar trends as the WER results. The knowledge learned from one dysarthric speaker tends to transfer to other speakers with similar IS. For example, Domotica speakers with IS below 65 get the best performance when pre-trained on the full Copas data which includes the speakers with IS below 60, while Domotica speakers with IS from 65 to 72 get better results when pre-trained on data exclude the part with IS $< 60$. For speakers with IS above 73, the acoustic model pre-trained with data with IS above 70 does the best. In our application, since collecting moderate impaired (with IS above 70) or normal speech needs less efforts than collecting severe dysarthric speech, we would consider pre-training on IS $> 70$ as the most beneficial choice which achieves fairly good improvements in most of cases without suffering any degradation dealing with speech from all impairment severities.

Besides that, in Figure 3 (a), the WER for dysarthric speech is mostly above 35\% and up to 80\%. Even with moderately impaired speech (IS ranges 70 to 75), the WERs are above 50\% in general. Therefore, the conventional pipeline structure with separate ASR and NLU modules cannot be applied to the dysarthric SLU task since errors generated during the ASR procedure would inevitably undermine the upstream NLU, which confirms our approach to extract knowledge from pre-trained ASR task and apply it to the user-taught E2E SLU task.

\begin{figure}[t]
  \setlength{\belowcaptionskip}{-0.3cm}
  \centering
  \includegraphics[width=\linewidth]{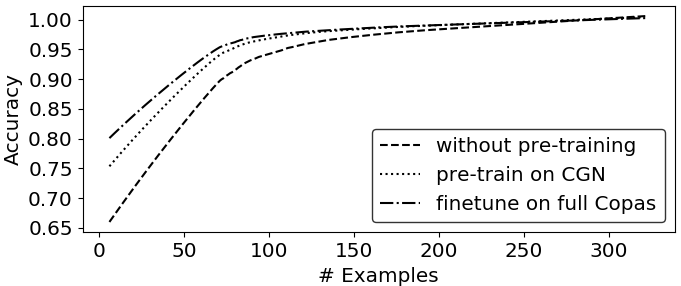}
  \caption{Learning curve for the Domotica corpus.}
  \label{fig2}
\end{figure}  
\begin{figure}[h]
  \setlength{\abovecaptionskip}{0.cm}
  \setlength{\belowcaptionskip}{-0.5cm}
  \centering
  \subfigure(a){\includegraphics[width=0.42\linewidth]{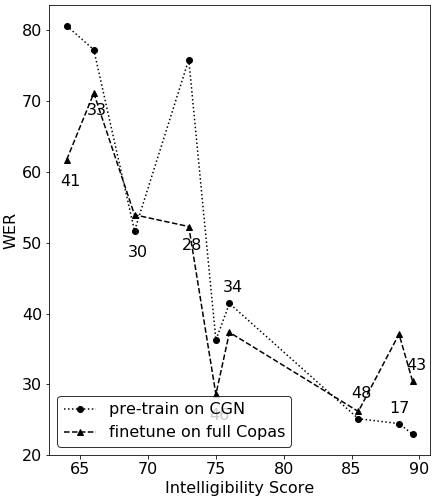}}
  \subfigure(c){\includegraphics[width=0.48\linewidth]{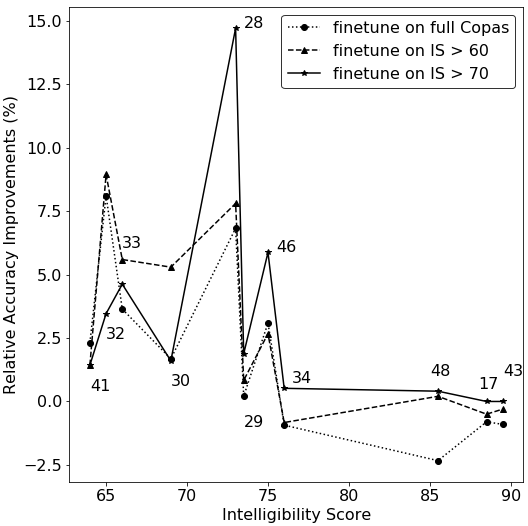}}
  \subfigure(b){\includegraphics[width=0.94\linewidth]{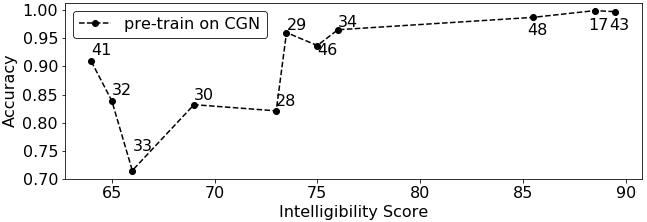}}
  \caption{Per speaker (a) WER (in \%) with and without finetuning on dysarthric data; (b) SLU accuracy with “pre- train on CGN”; (c) relative improvements compared with “pre-train on CGN” on the 15\% task-specific data sorted by intelligibility score.}
  \label{fig3}
\end{figure}

\section{Conclusions}
In this paper, we design a SLU system for dysarthric speech and investigate to which extent the dysarthric SLU task can benefit from pre-training with ASR targets on dysarthric speech.

The designed SLU system consists of a 16-layer TDNN based acoustic model which encodes the input features to the high-level bottleneck features and a 2-layer capsule network which decodes the bottleneck features to the intent slots. The acoustic model is pre-trained with ASR targets in two stages. We firstly construct an initial acoustic model trained on a large corpus of normal speech to learn the general knowledge, and then finetune the initial model with the mixture of dysarthric and normal speech corpus to model the distribution of dysarthric speech.

The designed SLU system is verified on a public Dutch dysarthric dataset. The performance gains reach up to 15\% absolute in terms of slot F1-score compared with the previous state-of-the-art model without pre-training. Average gains up to 5\% are found with respect to pre-training on normal speech, showing our pre-training strategies work. By introducing the IS to quantize impairment severity and comparing pre-training on utterances belonging to different severity levels, we conclude it is wise to adapt the models with speech of similar impairment severity levels, in order to avoid degradation.
Finally, unlike the ASR task which fails miserably without pre-training on dysarthric data, the user-taught SLU approach still reaches viable accuracies without pre-training or with pre-training on normal speech. Hence omitting dysarthric data collection might be an option in some deployments, though a price needs to be paid in terms of learning speed.
%\textcolor{red}{\sout{However, gains are limited, so unlike the ASR task, in a user-taught SLU system, no pretraining or pretraining on normal speech are suboptimal, but reasonable options.}=$>$However, gains are limited. Since a user-taught SLU system requires user efforts, collecting data from speech impaired users is not as efficient as from healthy users. So unlike the ASR task, in this application, pre-training on normal speech are suboptimal, but reasonable options when taking users’ efforts into consideration. (can we remove the phrase "no pre-training" here? Because in figure.2 as well as the first paragraph of section 4 (especially the last sentence), we indicate the improvement by using pre-training. If we think the performance falls caused by no pre-training could be tolerated, it will be strange that we take the fact we could get higher accuracy with less training samples as a strength to point out)}

\section{Acknowledgements}
The research was supported by the program of China Scholarship Council No. 201906090275, KUL grant CELSA/18/027 and the Flemish Government under “Onderzoeksprogramma AI Vlaanderen”.

\bibliographystyle{IEEEtran}
\normalem
\bibliography{main}

\end{document}